\documentclass{jfm}
\usepackage{graphicx}
\usepackage{epstopdf, epsfig}
\usepackage{amsmath}
\usepackage{mathtools}
\usepackage{natbib}
\usepackage{array}
\newcolumntype{C}[1]{>{\centering\let\newline\\\arraybackslash\hspace{0pt}}m{#1}}

\shorttitle{Bandwidth, spectral shape, and wave breaking onset}
\shortauthor{M.L. McAllister et al.}

\title{The influence of spectral bandwidth and shape on deep-water wave breaking onset}

\author{M.L. McAllister\aff{1}, N. Pizzo\aff{2}, S. Draycott\aff{3} \& T.S. van den Bremer\aff{1,4} }

\affiliation{\aff{1}Department of Engineering Science, University of Oxford, Oxford OX1 3PJ, UK
\aff{2} Scripps Institution of Oceanography, University of California, San Diego, La Jolla, California 92037, USA
\aff{3} School of Engineering, University of Manchester,
Manchester M60 1QD, UK
\aff{4} Faculty of Civil Engineering and Geosciences, Delft University of Technology, 2628 CD, Delft, The Netherlands}

\begin{document}
	
	\maketitle
	
	\begin{abstract}
Deep-water surface wave breaking affects the transfer of mass, momentum, energy and heat between the air and sea. Understanding when and how the onset of wave breaking will occur remains a challenge. The mechanisms that form steep waves, i.e. nonlinearity or dispersion, are thought to have a strong influence on the onset of wave breaking. In two-dimensions and on deep-water, spectral bandwidth is the main factor that affects the roles these mechanism play. Existing studies, in which the relationship between spectral bandwidth and wave breaking onset is investigated, present varied and sometimes conflicting results. We perform numerical simulations of two-dimensional focused wave groups on deep-water to better understand this relationship, with the aim reconciling existing studies. We show that the way in which steepness is defined, may be the main source confusion in the literature. At breaking onset, locally defined steepness reduces as a function of bandwidth, and globally defined steepness increases.
 The relationship between global breaking onset steepness and spectral shape (using the parameters bandwidth and spectral skewness) is too complex to parameterise in a general sense. However, we find that the local surface slope of maximally steep non breaking waves, of all spectral bandwidths and shapes (constant-steepness, constant-amplitude, and JONSWAP), approaches a limit of $1/\tan(\pi/3)\approx0.5774$. This slope based threshold, is simple to measure and may be used as an alternative to existing kinematic breaking onset thresholds. There is a potential link between slope based and kinematic breaking onset thresholds which future work should seek to better understand.
 
 
\end{abstract}
	
\section{Introduction}\label{sec:Intro}
Wave breaking poses an upper limit to how steep an individual wave can become, is the main mechanism of dissipation of wave energy in the ocean and determines how sea states evolve \citep{hasselmann1974,phillips1985, young2006,iafrati2011,romero2012,sutherland2013}. Wave breaking also affects turbulence and mixing in the upper ocean and contributes significantly to air-sea interaction \citep{melville1996,deike2022}. The interaction of the atmosphere and the sea plays a leading role in the uptake of anthropogenic CO$_2$ by the ocean, and is thus of crucial importance to understanding climate change \citep{smith1985,reichl2020}. Understanding how and when waves break is, therefore, essential for forecasting (extreme) waves, predicting the resulting loads they exert on offshore structures, and climate modelling.

\begin{figure}
    \centering
    \includegraphics[width=0.6\textwidth]{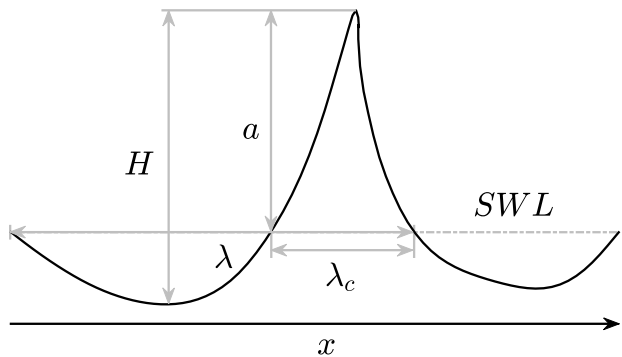}
    \caption{Definition of zero-crossing wave height $H$, wavelength $\lambda$ ($k=2\pi/\lambda$), crest length $\lambda_c$, and (crest) amplitude $a$ relative to the still-water level (SWL).}
    \label{fig:WaveProps}
\end{figure}

\cite{stokes1880} first proposed a limiting form for two-dimensional (2D) progressive waves on deep water, where the steepest possible wave crest encloses an angle of $120^\circ$. This waveform corresponds to a steepness of $kH/2=0.44$, where $H$ is wave height, and $k$ is wavenumber (see Fig. \ref{fig:WaveProps}). However, waves in the ocean are not of permanent progressive form and field and laboratory observations of waves steeper than this limit are not uncommon \citep{toffoli_etal2010a}, and observations of breaking below this threshold are also frequently made \citep{perlin2013}. Various external factors may cause deviations from this theoretical limit; for example the interaction of winds and currents with waves can strongly modulate the onset of breaking \citep{wu2004,babanin_etal2010,vrecica2022}. In the absence of external factors, the mechanism by which large wave crests are formed is also thought to influence breaking onset; waves that form as a result of (linear) dispersive focusing and those that from as a result of (nonlinear) modulational instability have been shown to break at different steepness \citep{perlin2013}. 

Broadly speaking, the different mechanisms (dispersion and nolinearity) that generate extreme (steep) waves crest are strongly influenced by the bandwidth of the underlying spectrum. In 2D, bandwidth corresponds to the range of frequencies over which wave energy is present; in 3D, bandwidth also corresponds to the degree of directional spreading of a given sea state. In moderately spread uni-modal conditions, directional spreading (directional bandwidth) has been shown to increase the steepness at which breaking onset occurs \citep{johannessen_swan2001,latheef2013}. For crossing-sea conditions, where the directional spectrum is bimodal, breaking onset steepness increases with the angle of crossing \citep{mcallister2019}. In general, increasing directional bandwidth appears to increases the steepness at which wave breaking onset occurs. It is worth noting that numerical studies involving `short-crested' waves created by varying the angle between two monochromatic waves demonstrate a non-monotonic increase in breaking onset steepness as the angle is increased \citep{roberts_schwartz1983,tsai1994}. Owing to the challenges associated with modelling experimentally and numerically highly directionally spread waves, the relationship between spreading and breaking onset is not fully understood.

\begin{figure}
    \centering
    \includegraphics[scale=0.6]{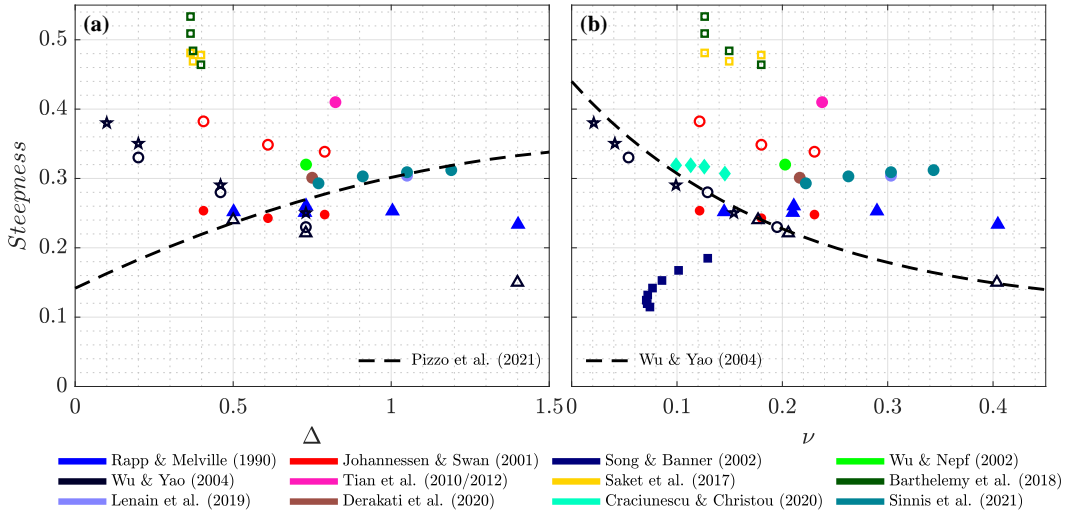}
    \caption{Breaking threshold steepness as a function of bandwidth taken from existing experimental and numerical studies. Filled markers show global (i.e., $S$) and open markers show local  measures of steepness (i.e., $ak$ or $kH/2$); circular markers correspond wave groups based on constant-steepness spectra, stars to linear steepness spectra, triangles to constant-amplitude spectra, diamonds to JONSWAP spectra, and squares to chirped wave packets (see Tab. \ref{tab:breaking studies}). Colours denote the different studies as given in the legend. Dashed lines show the parametric breaking thresholds \eqref{eq:Thresh_Pizz} and \eqref{eq:Thresh_Wu}.}
    \label{fig:previos_breaking}
\end{figure}

 In 2D, frequency bandwidth affects the duration over which waves may interact nonlinearly, meaning that modulational instability dominates when the underlying spectrum is narrow and waves have sufficient time to interact. Paired with a measure of steepness, bandwidth may be used to characterise the impact of nonlinearity versus dispersion in the steepening, and potential overturning, of surface waves \citep{pizzo2019}. However, there may not be a unique way to do this \citep{perlin2013} and
existing studies carried out in 2D appear not to reach clear consensus regarding frequency bandwidths effect on breaking onset (see Tab. \ref{tab:breaking studies} and Fig. \ref{fig:previos_breaking} for an overview of these studies). One example of this lack of consensus can be found when comparing \cite{wu2004} and \cite{pizzo2021}, who examine wave groups based on constant-steepness spectra created experimentally and numerically, respectively. \cite{wu2004} observe a decreasing relationship,
\begin{equation}\label{eq:Thresh_Wu}
kH/2=0.44e^{3.0\nu^2-3.9\nu}
\quad \textrm{for}\quad 0.021\leq\nu \leq0.404,
\end{equation}
(note in their Eq. (9) \cite{wu2004} use $a$ to denote $H/2$ in contrast to our notation defined in Fig. \ref{fig:WaveProps}), whereas \cite{pizzo2021} observe an increasing relationship,
\begin{equation}\label{eq:Thresh_Pizz}
S=-0.0579\Delta^2+0.2177\Delta+0.1417
\quad \textrm{for}\quad 0.2\leq\Delta \leq 1.6,
\end{equation} 
between the breaking onset steepness measures ($kH/2$, $S$) and the bandwidth measures ($\nu$, $\Delta$) these authors consider. A number of factors, such as inconsistent definitions of steepness, may explain this apparent contradiction. Furthermore, it is important to note that different studies have used a range of different spectral shapes to create breaking wave groups, such as constant-amplitude spectra \citep{rapp1990,wu2004}, constant-steepness spectra \citep{wu2004,pizzo2021}, JONSWAP spectra \citep{craciunescu2020}, and chirped wave packets \citep{song2002,saket2017,barthelemy2018,pizzo2019a}. 

When different spectral shapes are used, the definition of spectral bandwidth itself may become a source of inconsistency. For constant-amplitude and constant-steepness spectra, it is intuitive to define bandwidth as the range of frequencies over which the spectrum is defined $\Delta f$, and when using a JONSWAP spectrum the peak enhancement factor $\gamma$ may be used to define bandwidth. The parameter $\nu$, which is calculated as $\sqrt{m_0m_2/m_1^2-1}$ with $m_n$ the $n^{\rm th}$ spectral moment of the energy spectrum $E(f)$ 
, may be used to provide a more general definition of bandwidth. When $\nu$ is used as a measure of bandwidth, the results of \cite{craciunescu2020} (JONSWAP spectrum) show an opposite relationship of breaking onset steepness with bandwidth compared to \cite{pizzo2021} (constant-steepness spectrum) when the same measure of steepness is used as a breaking onset threshold, 
with the former \citep{craciunescu2020} decreasing in Fig. \ref{fig:previos_breaking}b and the latter \citep{pizzo2021} increasing in Fig. \ref{fig:previos_breaking}a (we note that the variation in breaking onset steepness observed in \cite{craciunescu2020} is very small, see Fig. \ref{fig:previos_breaking}b). This may suggest that, even if bandwidth is defined consistently, spectral shape has an influence on the relationship between bandwidth and breaking onset. 

Table \ref{tab:breaking studies} lists the breaking onset steepness threshold values found in comparable studies of 2D wave breaking onset (on deep and intermediate water depths). Figure \ref{fig:previos_breaking} plots breaking onset steepness reported in the studies listed in Tab. \ref{tab:breaking studies} as a function of the bandwidth measures $\Delta=\Delta f/f_0$ when applicable (left), where $f_0$ is the central frequency, and $\nu$ (right). We exclude studies that examine breaking onset of modulated wave trains from this figure, and note that modulated wave trains, given sufficient distance to propagate, can break with initial monochromatic steepness as low as $0.08$ \citep{babanin_etal2010}. Open markers denote local measures of steepness (e.g., $kH/2$, $ak$), and filled markers denote global steepness in the form of linearly predicted maximum surface slope ($S=\sum a_n k_n \equiv k_c \sum a_n\equiv a_0 k_c$, where $k_c$ is the characteristic wavenumber). 

The results of \cite{wu2004} show a significant inverse relationship between bandwidth and breaking onset steepness and were used to fit \eqref{eq:Thresh_Wu} \citep{wu2004}; their values of breaking onset steepness are measured locally. Local steepness is also reported in \cite{johannessen_swan2001}, \cite{saket2017}, and \cite{barthelemy2018}; all three studies report significantly higher breaking onset steepness than \cite{wu2004}. We note that \cite{saket2017} and \cite{barthelemy2018} report local crest front steepness $S_c=\pi a/\lambda_c$, where $\lambda_c$ is the crest length (see Fig. \ref{fig:WaveProps}). The values reported in \cite{johannessen_swan2001} decrease as a function of bandwidth, in a similar manner to \eqref{eq:Thresh_Wu} albeit with higher values of breaking onset steepness. In \cite{saket2017} and \cite{barthelemy2018} bandwidth does not vary sufficiently to observe a trend. \cite{johannessen_swan2001} also report values of global breaking onset steepness for the same experiments (solid markers of the same colour), which do not vary significantly with bandwidth; this is also the case for values of global steepness reported by \cite{rapp1990}. The numerical numerical results presented in \cite{song2002} show increasing global breaking onset steepness with bandwidth, over the small range of bandwidths that they performed simulations for. The experimental results reported in \cite{sinnis2021} show good agreement with the parameterisation \eqref{eq:Thresh_Pizz} (emphasizing the latter is obtained from numerical simulations). In \cite{sinnis2021}, the linear predictions of maximum surface slope (global steepness) are calculated using phase and amplitude measured close to the point of wave generation, which makes these results less reliant on perfect wave generation and most likely more accurate.  

The data as a whole do not show a clear and consistent relationship between breaking onset steepness and either measure of bandwidth. The majority of the data in Fig. \ref{fig:previos_breaking} are based on experiments in which constant-steepness spectra were used, so, while some of the scatter may be a result of the different spectral shapes studied, the scatter may not be explained by spectral shape alone. What is made clear in Fig. \ref{fig:previos_breaking}, and in \cite{rapp1990} and \cite{johannessen_swan2001}, is that using global and local definitions of steepness may lead to conflicting outcomes; 
we will investigate this in detail in \S \ref{sec:Lin_kin_Steep} and \S\ref{sec:Numerics}. 

Generating breaking waves experimentally is challenging, doing so over a wide range of bandwidths is even more so as experiments involving narrow bandwidths require very long flumes. Additionally, measuring the relevant spatial characteristics and kinematics of breaking waves experimentally is also difficult. Numerical simulations offer the ability to examine wave properties readily with high spatial resolution, although it is evidently more difficult to include in numerical simulations the non-potential-flow effects that start to occur after breaking onset. In this paper, we perform a series of 2D fully nonlinear Lagrangian potential-flow simulations based on the numerical method proposed in \cite{dold1986}, with the aim of elucidating the effects bandwidth and spectral shape have on wave breaking onset, and to reconcile existing studies, which appear to disagree. We analyse focused wave groups, varying spectral shape and bandwidth, and investigate how the onset of wave breaking is affected. We detail how we define focused wave groups in \S\ref{sec:wave_def}, we then examine properties of steep and breaking focused wave groups subject to linear (\S\ref{sec:Lin_kin_Steep}) and fully nonlinear evolution (\S\ref{sec:Numerics}). Finally, we discuss our results and draw conclusions in \S\ref{sec:conclusions}. 

\begin{table}
	\caption{Summary of comparable 2D laboratory (L) and numerical (N) studies of breaking onset on deep and intermediate water. Some of the values not presented explicitly in the cited manuscripts have been extracted from digitised figures, and several are taken from \cite{perlin2013}. Initial monochromatic steepness (IMS) is used to characterise the steepness of modulated wave trains (MWTs). For focused wave groups we present $S$ where available; values that correspond to local measures of steepness $kH/2$, $ak$, and $S_c$ are labelled with $*$, $\dagger$, and $\ddagger$ symbols, respectively. Experiments in which values of $S$ based on measurements were reported (i.e., $S=\sum a_nk_n \cos(\theta_n)$) we denote with a $\star$ symbol. Values of bandwidth $\nu$ that we have calculated based on the reported spectral shape are labelled with a letter $c$. For the chirped wave packets in \cite{pizzo2019a}, we report the 3-decibel limits (3dB) as a measure of bandwidth.}\label{tab:breaking studies}
	\begin{center}
		\begin{tabular}{lC{2.cm}C{3cm}C{3cm}}
			 & Spectrum & Bandwidth ($\nu$) &  Threshold steepness\\
			 \textbf{Modulated wave trains (MWTs):} & & & IMS ($a_0k_0$) \\
			 \hline
            \cite{tulin1999} - L& MWT & - & $0.22$-$0.41$ \\
            \cite{babanin_etal2010} - N & MWT & - &  $0.08$ \\
             &  &  &  $0.40^*$ \\

             \cite{tian2012} - L & MWT & - & $0.12$ \\
             \\
             \textbf{Constant-amplitude spectra:} & & & $S$ \\
			\hline  
			\cite{rapp1990} - L & $a_n=C$ & $0.150^c,0.210^c,0.211^c$ $0.290^c,0.405^c$& $0.252,0.251,0.260$ $0.253,0.234$ \\
		    \cite{chaplin1996} - L  &  $a_n=C$ & $-$ & $0.265$ \\
		     
		    		  \cite{wu2004} - L &  $a_n=C$ & $ 0.177,0.206,0.404$ & $0.240^{*},0.221^{*},0.150^{*}$ \\
		    		  \\
		    \textbf{Linear steepness spectra:} & & & $S$ \\
			\hline
			\cite{wu2004} - L & $a_nk_n=\frac{k_N-k_n}{k_N-k_1}C$ & $0.021,0.041,0.099$ $0.154$& $  0.380^{*},0.350^{*},0.290^{*}$ $0.250^{*}$ \\
		    \textbf{Constant-steepness spectra:} & & & $S$ \\
            \hline
          \cite{chaplin1996} - L       & $a_nk_n=C$ & $-$ & $0.300$ \\
		  \cite{johannessen_swan2001} - L & $a_nf^{-2}_n=C$  &$0.121^\mathrm{c} ,0.180^\mathrm{c},0.230^\mathrm{c}$& $0.253,0.243,0.248$\\  
		  &&&$0.338^{\dagger},0.348^{\dagger},0.382^{\dagger}$\\
		  \cite{wu2002} - L & $a_nk_n=C$ & $ 0.211^\mathrm{c}$ & $0.320$ \\
		  \cite{wu2004} - L &  $a_nk_n=C$ & $0.054,0.129,0.195$ & $0.330^{*},0.280^{*},0.230^{*}$ \\
		  \cite{tian2012} - L &$a_nk_n=C$ & $0.238^\mathrm{c}$ &   $0.41$\\ 
		  \cite{lenain2019} - L & $a_nk_n=C$& $0.303^\mathrm{c}$ & $0.304^\star$\\
            \cite{derakhti2020} - N & $a_nk_n=C$& $0.217^\mathrm{c}$ & $0.301^\star$\\
		  \cite{sinnis2021} - L& $a_nk_n=C$& $0.222^\mathrm{c},0.263^\mathrm{c},0.303^\mathrm{c}$ $0.344^\mathrm{c}$ & $0.293^\star,0.303^\star,0.309^\star$ $0.312^\star$\\
          \cite{pizzo2021} - N & $a_nk_n=C$ & $0.058^\mathrm{c}$-$0.462^\mathrm{c}$ &$0.1737$-$0.3974$\\
          \\
            \textbf{Chirped wave packets (CWPs):}  & & & $S$ \\
           \hline
           \cite{song2002} - N & CWP & $0.074^\mathrm{c}$-$0.129^\mathrm{c}$&$0.115$-$0.185$\\
           \cite{saket2017} - L & CWP & $0.126^\mathrm{c},0.150^\mathrm{c},0.180^\mathrm{c}$ &$0.481^{\ddagger},0.469^{\ddagger},0.478^{\ddagger}$ \\
		   \cite{barthelemy2018} - N & CWP & $0.126^\mathrm{c},0.126^\mathrm{c},0.150^\mathrm{c}$ $0.180^\mathrm{c}$ &$0.534^{\ddagger},0.509^{\ddagger},0.484^{\ddagger}$ $0.464^{\ddagger}$ \\
		\cite{pizzo2019a} - N & CWP & 3dB: $0.2$-$0.5$  & $0.160$-$0.220$\\
\\
             \textbf{JONSWAP spectra:} & & & $S$ \\
		   \hline 
		   \cite{craciunescu2020} - L & JONSWAP& $0.099^\mathrm{c},0.113^\mathrm{c},0.126^\mathrm{c}$ $0.145^\mathrm{c}$ & $0.319, 0.319,0.317$ $0.307$ \\

		\end{tabular}
	\end{center}
\end{table}
\section{Definitions of focused wave groups}\label{sec:wave_def}
In the following sections (\S\ref{sec:Lin_kin_Steep} and \S\ref{sec:Numerics}), we will examine properties of focused wave groups relevant to wave breaking, first examining results based on linear wave theory in \S\ref{sec:Lin_kin_Steep} and then based on fully nonlinear numerical simulations in \S\ref{sec:Numerics}. In both cases, initial conditions are obtained using linear wave theory in the same manner, which we define in this section. 

\subsection{Linear initial conditions}
We define linear surface elevation
\begin{equation}
\eta^{(1)}(x,t)=\sum_{n=1}^{N}a_{n}\cos(\theta_n),\quad \mathrm{where}\quad \theta_n=k_n x-\omega_n t+\varphi_n,
\label{eq:eta_1_initial}
\end{equation}
and velocity potential
\begin{equation}
\phi^{(1)}(x,z,t)=\sum_{n=1}^{N}a_{n}\frac{\omega_n}{k_n}\exp\left(k_n z\right)\sin(\theta_n),
\label{eq:phi_1_initial}
\end{equation}
as a summation of $N$ free wave components propagating on deep water, with amplitude $a_n$, frequency $\omega_n$, and wavenumber $k_n$, obeying the linear dispersion relationship, $\omega_n^2=gk_n$ with $g$ the gravitational acceleration. Waves propagate in the positive $x$-direction, $z$ is positive in the upwards direction, with $z=0$ corresponding to the still-water level, and $t$ is time. The phases $\theta_n$ are defined such that all components are in phase at the desired focus time ($t=0$) and position ($x=0$). Defining phase in this manner creates a focused wave group, assuming linear dispersive focusing.
\subsection{Frequency spectra}
In the existing literature, various different spectra have been used to generated breaking wave groups (see Tab. \ref{tab:breaking studies}). To understand why differences in breaking behaviour observed in previous studies may arise, we will examine three spectral shapes, namely constant-steepness and constant-amplitude spectra (\S \ref{sec:Constant-amplitude and constant-steepness spectra}), and JONSWAP spectra (\S \ref{sec:JONSWAP spectra}). In each case, we will define the amplitude spectrum, $\hat{\eta}^{(1)}(f)=\int\eta^{(1)}(t)\exp(i2\pi ft){\rm d}t$, from which initial conditions \eqref{eq:eta_1_initial}-\eqref{eq:phi_1_initial} are obtained using properties of linear dispersion.

%
\subsubsection{Constant-amplitude and constant-steepness spectra}
\label{sec:Constant-amplitude and constant-steepness spectra}
Perhaps the most simple and the most commonly used (cf. Tab. \ref{tab:breaking studies}) spectra are those for which the spectral components have constant amplitude or steepness, for which the amplitude spectra have the general form: 
\begin{equation}\label{eq:fn_spec}
    \hat{\eta}^{(1)}(f)\propto f^m\qquad \mathrm{for} \qquad f_0-\frac{\Delta f}{2} \leq f \leq f_0+\frac{\Delta f}{2},
\end{equation}
where $f_0$ is the central frequency, and $\Delta f$ is the range of frequencies over which spectral components are defined (made non-dimensional as $\Delta=\Delta f/f_0$). For constant-amplitude $a_n=C$ ($m=0$ in \eqref{eq:fn_spec}) or constant-steepness $a_nk_n=C$ ($m=-2$ in \eqref{eq:fn_spec} in deep water) spectra, the bandwidth parameter $\Delta f$ has a well-defined effect on the distribution of wave amplitude (and energy) as a function of frequency. For both spectral shapes, it is possible to define spectra with a wide range of bandwidths ($0<\Delta<2$). 
 
\subsubsection{JONSWAP spectra}
\label{sec:JONSWAP spectra}
A drawback of using constant-amplitude and constant-steepness spectra is that such simple spectra don't represent well the complete spectral shape of realistic ocean waves. A JONSWAP spectrum,
\begin{equation}\label{eq:JONSWAP}
    \hat{\eta}^{(1)}(f)\propto g^2(2\pi)^{-4}f^{-5}\exp\left(-\frac{5}{4}\left(\frac{f}{f_p}\right)^{-4}\right)\gamma^\beta
    \quad \textrm{with}\quad%
    \beta=\exp\left(\frac{-(f/f_p-1)^2}{2\sigma^2}\right),
\end{equation}
provides a more realistic alternative, from which breaking wave groups of different bandwidths may be generated by varying the peak enhancement parameter $\gamma$. In \eqref{eq:JONSWAP},  $f_p$ is the peak frequency of the spectrum, and $\sigma$ takes the values $0.07$ and $0.09$ when $f<f_p$ and $f>f_p$, respectively.  The parametric form of the JONSWAP spectrum in \eqref{eq:JONSWAP} generally corresponds to an energy spectrum (and thus $a_n\propto\sqrt{E(f_n)}$). Instead, we set the amplitude spectrum to be proportional to the JONSWAP spectrum itself in \eqref{eq:JONSWAP}, as this gives the correct shape of extreme (and thus breaking) waves in an underlying random Gaussian sea \citep{lindgren1970,boccotti1982}.


\subsection{Varying bandwidth for different spectral shapes}
\begin{figure}
 	\centering
 	\includegraphics[scale=0.6]{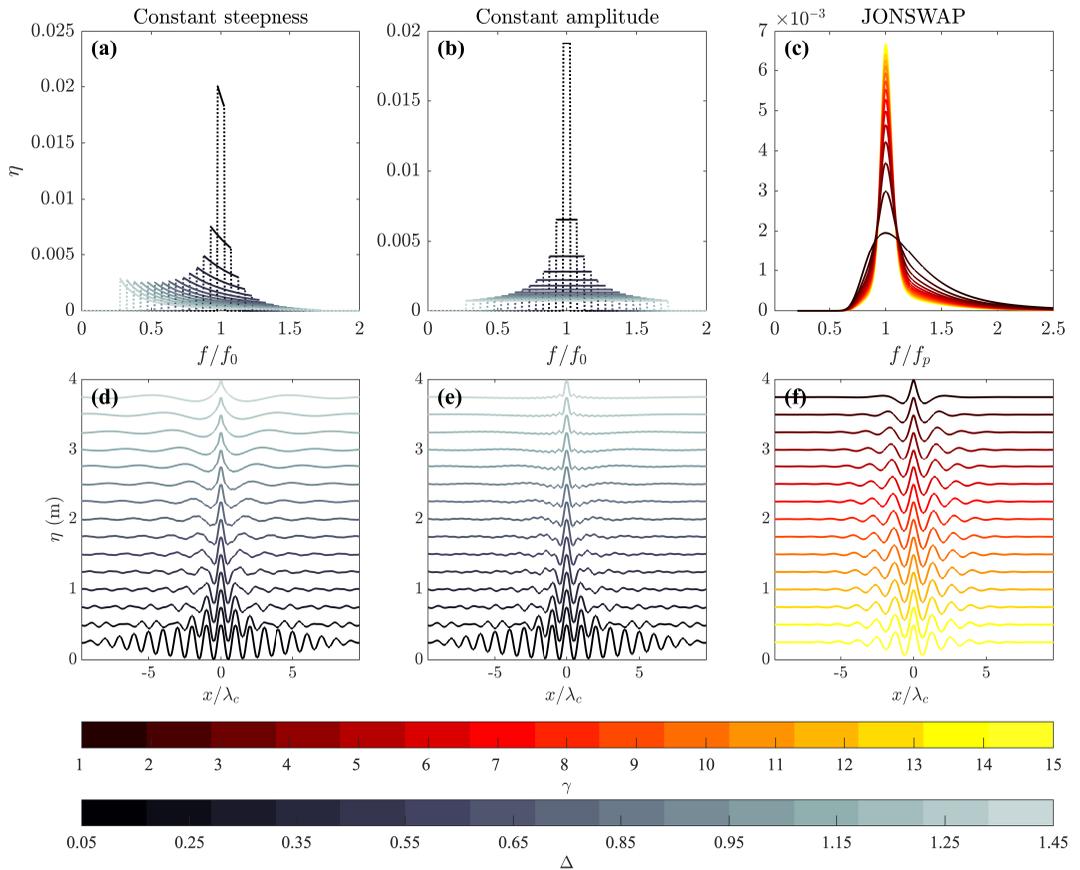}
 	\caption{Spectral shapes and corresponding linear focused wave groups used for calculations in Fig. \ref{fig:Spec_shapes_props_BW}; (a) and d) constant-amplitude, (b) and (e) constant-steepness, and (c) and (f) JONSWAP spectra. Bandwidth is varied using the parameters $\Delta f$ (panels a, b, d, and e) and $\gamma$ (panel c, and f); the colour scales denote the corresponding values of $\Delta$ (constant amplitude and constant steepness) and $\gamma$ (JONSWAP).}
 	\label{fig:Spec_shapes}
 \end{figure}
For all spectral shapes (i.e., \eqref{eq:fn_spec} and \eqref{eq:JONSWAP}), the amplitude of the spectrum $\hat{\eta}(f)$, $a_0$, is scaled to achieve the desired global steepness $S$ of the corresponding wave group. Varying bandwidth indirectly, using parameters such as $\Delta$, $\gamma$, and $\sigma_f$ for a given spectral shape, causes the corresponding mean or characteristic wavenumber,
\begin{equation}
    k_c=\frac{\sum a_n k_n}{\sum a_n},
    \label{eq:characteristic_wavenumber}
\end{equation}
to change value. Herein, when comparing wave groups of different bandwidths, we adjust the value of $f_0$ or $f_p$ to maintain a constant value of $k_c$. By keeping constant $k_c$, wave groups of equal $S$ are also of equal amplitude, as $S\equiv k_c\sum a_n$ with $a_0=\sum a_n$. For all wave groups we examine herein, we choose a characteristic frequency of $f_c=1$ Hz ($T_c=1$ s, $\omega_c=2\pi$ rad/s, and $k_c=(2\pi)^2/g$ rad/m).

Examples of the spectra and resulting focused wave groups we use for linear calculations are shown in Fig. \ref{fig:Spec_shapes}. The color scales denote the values of parameters $\Delta$ and $\gamma$ that correspond to the spectra and the wave groups in each panel; the same values and colours are also used in Fig. \ref{fig:Spec_shapes_props_BW}. We note that the parameters $\Delta$ and $\gamma$ are not direct measures of bandwidth, and varying $\Delta$ and $\gamma$ also changes other moments such as skewness. In the following sections, to provide a more general discussion on the role of bandwidth, we use the parameter $\nu$ unless stated otherwise, which is calculated as $\sqrt{m_0m_2/m_1^2-1}$ with $m_n$ the $n^{\rm th}$ spectral moment of the energy spectrum $E(f)$.

\section{Linear predictions of surface kinematics and slope}\label{sec:Lin_kin_Steep}
For waves with narrow-banded underlying spectra, nonlinear focusing brought about by third-order quasi-resonant wave-wave interactions (modulational instability) causes the formation of large waves that may eventually lead to breaking. Modulated wave trains with low initial monochromatic steepness (IMS, $a_0k_0>0.08$) may evolve to breaking given sufficient time \citep{peregrine1985,babanin_etal2010}. The local steepness, as waves approach breaking, asymptotes to a value of $kH/2=0.44$ for waves generated in this manner \citep{babanin_etal2010}. For waves with broad underlying spectra, however, the role of nonlinear focusing is reduced, and linear approximations may provided reasonable predictions of surface kinematics and slope.

In the following section we will use linear wave theory to calculate approximate values of surface kinematics and slope to better understand how bandwidth and spectral shape may affect breaking onset. For all the wave groups in this section, the same value of global wave steepness $S=1$ is maintained. We will revisit this with fully nonlinear simulations in \S \ref{sec:Numerics}.

\subsection{Surface kinematics}
  \begin{figure}
 	\centering
 	\includegraphics[scale=0.6]{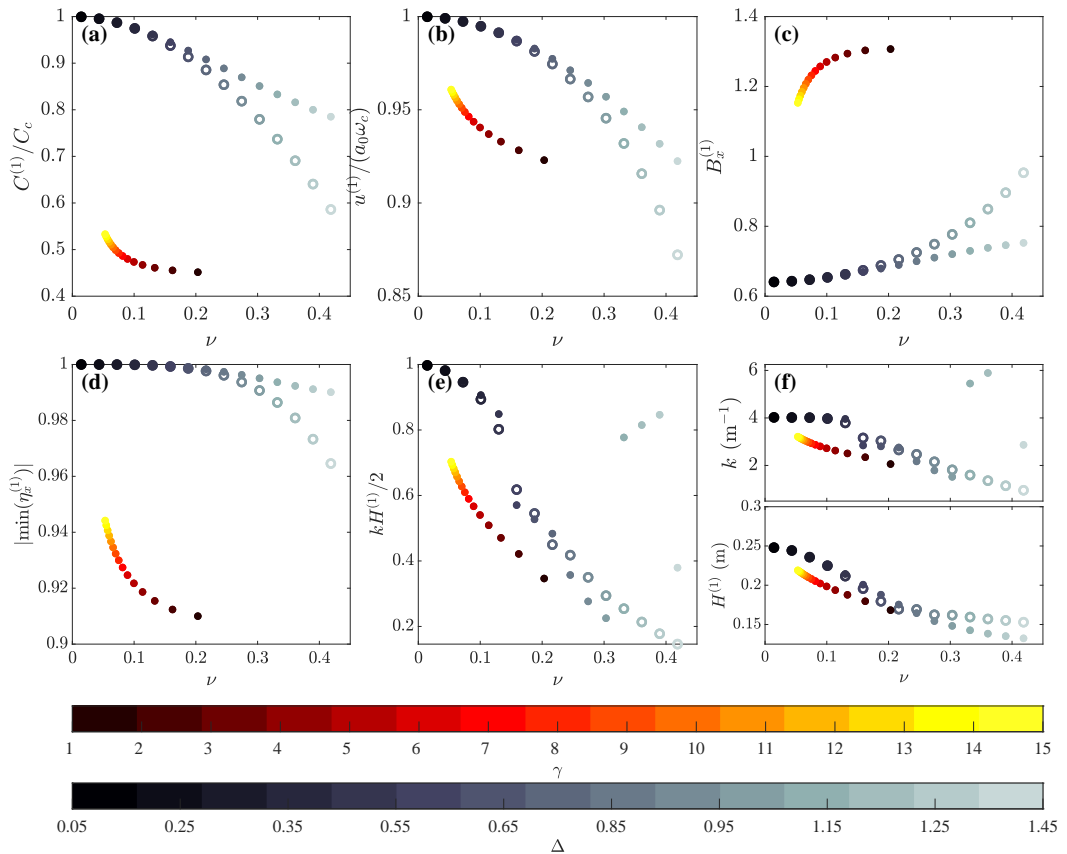}
 	\caption{Focused wave crest kinematics and measures of surface slope plotted as a function of bandwidth $\nu$, calculated based on linear theory at time and position of linear focus ($t=0$, $x=0$), for wave groups based on constant-amplitude (filled grey markers), constant-steepness (open grey markers), and JONSWAP spectra  (filled coloured markers). The colour scales denote the corresponding values of $\gamma$ (JONSWAP) and $\Delta$ (constant amplitude and constant steepness).}
 	\label{fig:Spec_shapes_props_BW}
 \end{figure}
The kinematic description of wave breaking, which makes use of the ratio of fluid velocity to crest speed, provides an intuitive means to explore the effects bandwidth and spectral shape may have on breaking onset. Following kinematic arguments, wave breaking occurs when the fluid velocity at a wave crest reaches or exceeds the crests velocity. The ratio of fluid velocity to crest speed is also the basis for the parameter $B_x$, which was derived using dynamical arguments in \cite{barthelemy2018}, and has been shown to predict the onset of breaking for 2D \citep{saket2017} and moderately spread waves \citep{barthelemy2018} in a range or water depths \citep{derakhti2020}. In these studies, breaking occurs when $B_x>0.855$. Therefore, we use the parameter $B_x$ to refer to the ratio of fluid to crest speed hereafter. In the following section we will use linear wave theory to examine how bandwidth affects wave crest kinematics and thus the ratio of fluid to crest speed $B_x$. 

\subsubsection{Crest speed}
The dispersive nature of surface gravity waves causes the apparent crest speed of an unsteady wave group to fluctuate as wave components of varying speeds interact coming in and out of phase \citep{fedele2020}. This effect has been observed in field data, and crest slow-down in particular has been linked to breaking onset \citep{banner2014}. Crest speed fluctuation is a predominantly linear effect that is also affected by nonlinear changes to dispersion as waves become steeper \citep{fedele2020}. The degree to which crest speed varies is related to the bandwidth of the underlying spectrum, as is the degree to which nonlinearity will affect dispersion. Two-dimensional wave groups with broad underlying spectra will experience greater linear crest speed variation and reduced nonlinear dispersion. 

If we define a crest as the point at which slope $\partial\eta/\partial x=0$, which occurs at $x=x_c(t)$, crest speed $C^{(1)}\equiv{\rm d}x_c/{\rm d} t$, where the superscript denotes this is a linear approximation. For a discrete spectrum of $N$ waves at linear focus ($x=0$, $t=0$), this may be expressed as,
\begin{equation}
C^{(1)}=\frac{\sum_{n=1}^{N}a_{n}\omega_nk_n}{\sum_{n=1}^{N}a_{n}k_n^2}.
\end{equation}

Linearly predicted crest speed, normalised by characteristic phase speed $C_c$, is shown as a function of bandwidth $\nu$ in Fig. \ref{fig:Spec_shapes_props_BW}(a) for the three different spectral shapes we examine. We use characteristic wavenumber (defined in \eqref{eq:characteristic_wavenumber}) and frequency to calculate phase speed: $C_c=\omega_c/k_c$. For all three spectral shapes, crest speed reduces with increasing bandwidth (n.b., the normalization by $C_c$ is independent of bandwidth, as we maintain a constant value of $k_c$). For the JONSWAP spectra, which extend over a smaller range of bandwidths $\nu$, the rate of crest slow-down is similar to the constant-amplitude spectra, but the normalised values of crest speed are much lower. This difference is a result of the high-frequency tail of the JONSWAP spectrum. 



\subsubsection{Fluid velocity}
Horizontal fluid velocity at $z=0$ and at focus may be calculated linearly  for a discrete spectrum of $N$ waves as
\begin{equation}
u^{(1)}=\frac{\partial\phi^{(1)}}{\partial x}\bigg|_{x=0, z=0, t=0}=\sum_{n=1}^{N}a_{n}\omega_n.
\end{equation}
Although evidently lower than the (nonlinear) velocity at the crest of a wave ($z=\eta$), linear fluid velocity at $z=0$ can still be used to illustrate how bandwidth affects surface kinematics. 
%
Linearly predicted horizontal fluid velocity, normalised by characteristic velocity $a_0 \omega_c$ (constant as $\nu$ is varied), are plotted as a function of bandwidth $\nu$ for the different spectral shapes we examine in Fig. \ref{fig:Spec_shapes_props_BW}(b). 
As bandwidth is increased, fluid velocity reduces in similar manner to crest speed but to a lesser extent. 


\subsubsection{Kinematic breaking parameter}
For all three spectral shapes  
$B_x^{(1)}\equiv u^{(1)}/C^{(1)}$ increases as a function of bandwidth, as shown in Fig. \ref{fig:Spec_shapes_props_BW}(c). This implies that for a certain value of global steepness $S$, which is kept constant as bandwidth is varied in our calculations, waves of broader bandwidth are more likely to break. In other words, linear wave theory suggests breaking will occur (if it occurs a fixed value of $B_x^{(1)}$) at lower steepness for wave groups with broader underlying spectra. 

\subsection{Steepness and wave slope}
For a discrete spectrum of $N$ waves, linearly surface slope is given by:
\begin{equation}
 \eta^{(1)}_x=\frac{\partial\eta^{(1)}}{\partial x}=\sum^N_{n=1}-a_n k_n \sin(\theta_n),\quad \mathrm{where} \quad \theta_n=k_n x+\omega_n t + \varphi_n.
\end{equation}
The maximum possible value of linearly surface slope, $\max (|\partial\eta^{(1)}/\partial x|)$, is thus
\begin{equation}
S=\sum_{n=1}^{N}a_{n}k_n\equiv k_c\sum_{n=1}^{N}a_{n}\equiv a_0 k_c,
\end{equation}
which is realised when the phases $\theta_n$ of all wave components are simultaneously $\pi/2$.
This global steepness $S$ is used to parameterize a range of breaking-related phenomena such as breaking intensity \citep{drazen2008}. \cite{pizzo2019a} and \cite{pizzo2021} demonstrate, when paired with $\Delta$, global steepness $S$ functions well as a parameter to predict breaking onset for chirped wave groups and wave groups based on a constant-steepness spectra (cf. \eqref{eq:Thresh_Pizz}). In these studies, the parameter $S$ functions as a global measure of nonlinearity (steepness) and $\Delta$ as measure of the degree to which dispersion or nonlinear focusing will occur (bandwidth). 

By definition, linear waves have the property $\mathrm{max}(|\eta^{(1)}_x|)=\mathrm{max}(\eta^{(1)}_x)=-\mathrm{min}(\eta^{(1)}_x)$, which is not necessarily true for nonlinear waves, for which $\mathrm{max}(\eta_x)\neq-\mathrm{min}(\eta_x)$. Wave breaking is initiated by overturning and crest instabilities that occur at the crest front, and hence when referring to maximum local steepness we report $|\mathrm{min}(\eta_x)|$. Nonlinearity, phase coherence, and imperfect wave generation (in the laboratory) may mean that actual wave slope at the point of wave breaking (local) will differ from a global measure such as $S$. To generate a (crest-) focused wave group, the phases of wave components are selected such that $k_nx-\omega_n t+\varphi_n=0$ at $t=0$ and $x=0$. To generate a maximum steepness wave group, the phases of all wave components $(k_nx-\omega_n t+\varphi_n)$ must be equal to $\pi/2$. These two conditions are evidently incompatible, except for monochromatic waves, and thus maximum local slope $|\mathrm{min}(\eta^{(1)}_x)|$ will only tend to $S$ as $\nu\rightarrow0$. In Fig. \ref{fig:Spec_shapes_props_BW}(d)   $|\mathrm{min}(\eta^{(1)}_x)|$ is plotted as a function of bandwidth. The difference between $|\mathrm{min}(\eta^{(1)}_x)|$ and $S$ (which we set to 1) increases with bandwidth and more significant for the JONSWAP spectra. For all three spectral shapes and all values of $\nu$, the difference between local ($|\mathrm{min}(\eta^{(1)}_x)|$) and global slope ($S$) is less than $10\%$. 

In \cite{wu2004} and many other studies, local steepness $kH/2$, where $H$ is wave height and $k$ is wavenumber (based on some local measure of period $T$ or wavelength $\lambda$), is used to examine wave breaking onset. \cite{wu2004} show that $kH/2$ functions well as parameter to predict the onset of wave breaking for wave groups based on constant-steepness and constant-amplitude spectra (cf. \eqref{eq:Thresh_Wu}). As introduced in \S \ref{sec:Intro}, the parameterisations \eqref{eq:Thresh_Pizz} and \eqref{eq:Thresh_Wu} (from \citet{pizzo2021} and \citet{wu2004}, respectively) describe opposing relationships between bandwidth and breaking steepness. In Fig. \ref{fig:Spec_shapes_props_BW}(e) linearly calculated local wave steepness $kH^{(1)}/2$ is plotted as a function of bandwidth; as bandwidth increases, $kH^{(1)}/2$ decreases for all three spectral shapes. Both wave local wavenumber $k$ and height $H^{(1)}$ decrease as a function of bandwidth (cf. Fig.\ref{fig:Spec_shapes_props_BW}(f) and (g), respectively). For constant-amplitude spectra the decreases in $kH^{(1)}/2$ is not smooth; this is a result of jumps in the position of zero-crossings that are used to calculate local wavenumber $k$ (Fig. \ref{fig:Spec_shapes_props_BW}(f)). Vertical asymmetry that arises from finite bandwidth, which causes $H^{(1)}$ to decrease, is well established linear effect (e.g., \cite{boccotti1982}). This local measure of steepness $kH^{(1)}/2$ varies by as much as $80\%$ as bandwidth is increased in Fig. \ref{fig:Spec_shapes_props_BW}(e), whereas local slope $|\mathrm{min}(\eta^{(1)}_x)|$ varies only by $4$-$10\%$ in Fig. \ref{fig:Spec_shapes_props_BW}(d) (for wave groups of constant $S$).  

 The reduction in local steepness with bandwidth we predict using linear wave theory can, at least partially, explain the behaviour of \eqref{eq:Thresh_Wu}, and thus the perceived conflicting relationships between breaking onset and bandwidth reported in \cite{pizzo2021} and \cite{wu2004}. We note that in our calculations $H^{(1)}$ and $k$ are calculated spatially, using zero-crossing methods, and in \cite{wu2004} the same values are calculated using time-domain measurements, and the linear dispersion relationship is used to calculate $k$ from the zero-crossing period. Even for linear calculations there is a degree of error associated with this method used in \cite{wu2004}. Additionally, in \cite{wu2004} measurements were only made at locations where a wave gauge was located, which is not necessarily the location of the maximum local steepness (this is an issues for all experiments where  wave gauge measurements are used).


\cite{rapp1990} also report that for experimentally generated incipient breaking waves, local steepness decreased with bandwidth and global steepness stayed approximately constant ($a_{sb}k_c$ and $ak_c$ in their notation, respectively). For these experiments focused wave groups were generated using constant-amplitude spectra; this different spectral shape may be a reason why $S$ remained approximately constant and did not increase with bandwidth as observed in \cite{pizzo2019}.


\section{Fully nonlinear numerical simulations}\label{sec:Numerics} 
In 2D, third-order quasi-resonant wave-wave interactions lead to modulation instability or nonlinear focusing when steepness $S$ is high and bandwidth is low. In such cases, nonlinear focusing may lead to large deviations from linear wave theory. 
In the following section, we perform numerical simulations using a fully nonlinear boundary integral method \citep{dold1986} to examine how bandwidth and spectral shape may influence breaking onset. The numerical model is Lagrangian and allows for the simulation of double-valued surfaces to the point of reconnection. Therefore, we can simulate breaking waves, and in what follows we will identify breaking onset to occur when the free surface becomes vertical. The numerical method has been widely used and validated in similar studies \citep[e.g.,][]{dold1992,henderson1999,song2002,pizzo2019a,pizzo2021}.

We first explain the numerical domain and initial conditions we use to perform simulations (\S\ref{sec:NumDomain}), we then investigate global and local measures of steepness using simulations based on constant-steepness spectra (\S \ref{sec:Steepness}), and finally investigate spectral shape (\S \ref{sec:spec_shape}) and breaking onset detection (\S \ref{sec:onset}) using simulations based on constant-steepness, constant-amplitude, and JONSWAP spectra.

\subsection{Numerical method}\label{sec:NumDomain}
The numerical method we use \citep{dold1986} solves Laplace's equation subject to the kinematic and dynamic boundary conditions for free surface gravity waves in deep water, using an approach based on the Cauchy theorem boundary integral. This approach describes the time evolution of Lagrangian surface particles and is computationally efficient compared to boundary integral methods that are solved in the real domain.   

\subsubsection{Domain setup}
In defining the scale of our numerical domain we use similar parameters to those used in \cite{pizzo2021}; these are summarised in Tab. \ref{tab:Numerical Setup}. The initial length of the domain, at $t_0$ (we use the subscript $0$ to indicate initial values), $X=93.7$ m, which corresponds to $60$ characteristic wavelengths ($\lambda_c=2\pi/k_c$). For the simulations we present the domain is discretised using $2048$ particles, giving an initial particle spacing $\Delta_{X_0}$ of $4.6$ cm, which corresponds to approximately $34$ particles per wavelength (Configuration A).  

For a number of simulations we apply a conformal map to redistribute particles so that they are clustered around the focused wave group one period prior to the time of wave breaking $t^\star$ 
 (Configuration B). The initial particle positions $x_p$ are projected onto a unit circle $\zeta=\exp(i2\pi x_p/X)$ and then re-mapped,
\begin{equation}\label{eq:con_map}
    \Theta=\frac{\zeta+\nu_0}{1+\nu_0\zeta}\quad \textrm{for}\quad -1<\nu_0\leq0,
\end{equation}
where $\nu_0$ defines the degree of clustering. For the simulations we re-run using this mapping a value of $\nu_0=-0.5$ was used. This re-mapping gives a minimum particle spacing of $1.5$ cm or approximately $105$ particles per wavelength.

\begin{table}
	\caption{Numerical domain setup.}\label{tab:Numerical Setup}
	\begin{center}
		\begin{tabular}{ccccccc}
			Configuration &\multicolumn{2}{c}{Domain length} & \multicolumn{3}{c}{Number of particles and spacing} & \multicolumn{1}{c}{Initial time}\\
			&$X$ (m) & $X/\lambda_c$ & $N_p$ & $\Delta_{X_0}$ (m) & $\lambda_c/\Delta_{X_0}$ & $t_0$ (s) \\
			\hline
			 A (All cases) & 93.7 & 60 & 2048 & 0.0458 & 34.1 & $-L/(\nu C_{gc})$\\
			B (Selected cases) &93.7 & 60 & 2048 & 0.0148-0.1369 & 11.41-105.5 & $t^\star-T_c$\\
		\end{tabular}
	\end{center}
\end{table}

\subsubsection{Initial conditions}\label{sec:Num_ICs}
The numerical domain is periodic in space and models the time evolution of Lagrangian particles located on the free surface. Simulations are initiated using initial conditions in the form of a potential at each initial particle location (at the free surface). Initial particle spacing in $x$ may be defined arbitrarily and need not be regular. The smallest particle spacing affects the model's high-wavenumber resolution. We define our initial particle spacing regularly as $\Delta_{X_0}=X/N_p$, where $N_p$ is the number of particles.

We define initial conditions using linear wave theory (see \S \ref{sec:wave_def}). To reduce errors associated with using these approximate initial conditions, we choose a start time $t_0$ that ensures the waves are dispersed and have low initial steepness.  The time required to achieve this varies with bandwidth; we thus define initial time as $t_0=-L/(\nu C_{gc})$ (akin to  \cite{pizzo2021}), where $C_{gc}=\omega_c/(2k_c)$ is the characteristic group velocity in deep water and $L=10$ m is an arbitrary distance. Defining our initial particle distribution as uniformly spaced in $x$ results in a wavenumber discretization of $k_n=n 2\pi/X$ for $n=1$ to $N_p/2$. Initial conditions are scaled to give the desired value of $S$ at linear focus (see also \S \ref{sec:wave_def}). For JONSWAP spectra the finite wavenumber support results in a slight truncation of the high-frequency tail (at approximately $17k_c$ or $4f_c$, which is $6f_p$). Each simulation is run for a total time of $T=1.25|t_0|$. The time step of the model is defined dynamically to maintain a specified order of accuracy.

\subsection{Global and local measures of steepness}\label{sec:Steepness}
In \S \ref{sec:Lin_kin_Steep} we have demonstrated that, for wave groups of constant global steepness $S$, local steepness at focus ($kH/2$) decreases with bandwidth. We now examine how significantly this linearly predicted phenomenon affects nonlinear breaking waves, by simulating incipient breaking wave groups based on constant-steepness spectra. To produce simulations of incipient breaking waves for a range of bandwidths, we search for the steepest non-breaking wave group at each value of bandwidth by varying the input value of $S$. We make use of \eqref{eq:Thresh_Pizz} to provide an initial guess of this value of $S$. We stop searching when the step size in $S$ is less than $5\times10^{-4}$. 
\begin{figure}
 	\centering
 	\includegraphics[scale=0.6]{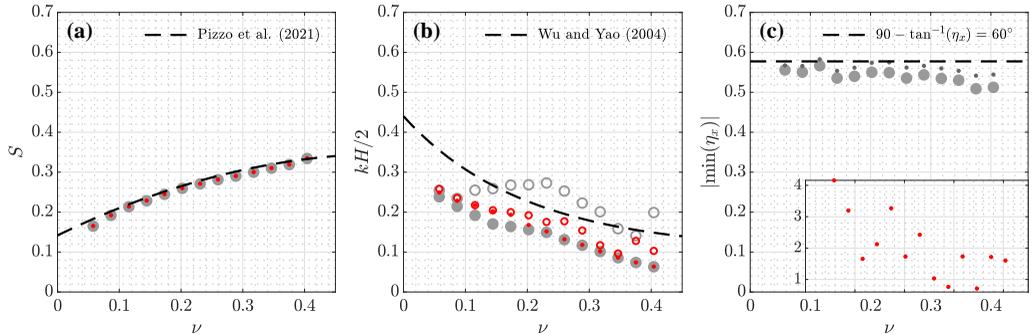}
 	\caption{Global and local measures of maximum steepness as a function of bandwidth for the steepest non-breaking (grey markers) and breaking (red dots) wave groups based on constant-steepness spectra. In panel (b) closed markers correspond to steepness measured $\pm T_c/2$ either side of the time of maximum slope $|\mathrm{min}(\eta_x)|$, and open markers correspond to maxima observed without this time constraint; in panel (c) the smaller dark grey dots correspond to simulations that were re-run using increased particle resolution at the crest (Config. B). The inset in panel (c) shows the (much higher) values of maximum slope of breaking wave groups.}
 	\label{fig:AK_local_global_steepness}
 \end{figure}
 
Figure \ref{fig:AK_local_global_steepness} shows global and local measures of steepness and slope for the largest non-breaking (grey large markers) and smallest breaking wave groups simulated (red small markers). In panel (a) global steepness $S$ is plotted as a function of $\nu$; the input global steepness $S$ of simulations involving incipient breaking waves follows the quadratic fit \eqref{eq:Thresh_Pizz} of \cite{pizzo2019}, as expected. The corresponding values of local steepness $kH/2$ are shown in Fig. \ref{fig:AK_local_global_steepness}(b), and are similar though not equivalent to the fit \eqref{eq:Thresh_Wu} of \cite{wu2004} with a clear offset between the two. The difference between the trends in our simulated results shown in panels (a) and (b) is consistent the conclusions we have already drawn through linear calculations in \S\ref{sec:Lin_kin_Steep}. Linear wave theory can therefore explain the difference between \eqref{eq:Thresh_Wu} and \eqref{eq:Thresh_Pizz} and help understand inconsistencies in the existing literature (cf. Fig. \ref{fig:previos_breaking}).

\begin{figure}
 	\centering
 	\includegraphics[scale=0.6]{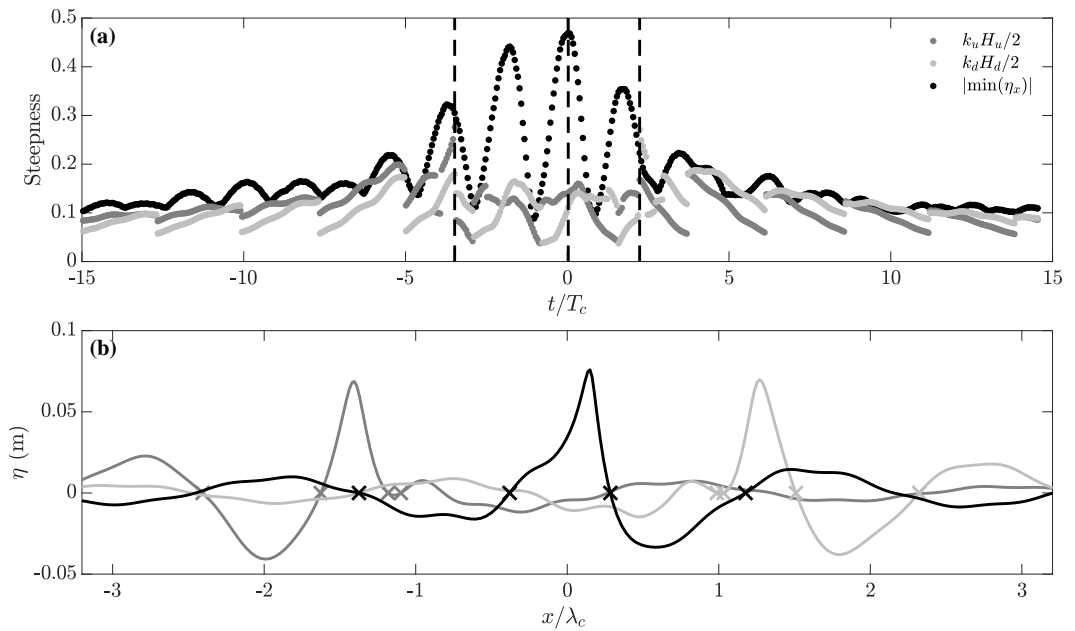}
 	\caption{Example of spikes in local steepness $kH/2$ calculated for a focusing wave group; panel (a) show up- ($k_u H_u/2$) and down- ($k_d H_d/2$) crossing steepness and maximum local slope of the largest wave as a function of time, panel (b) shows the surface elevations that correspond to maximum values of each parameter (of corresponding line colour) with zero crossings shown as $\times$ symbols. Vertical dashed lines in panel (a) show the three times at which maximum values of each parameter is observed, which correspond to the times for which surface elevations in space are shown in (b).}
 	\label{fig:HK_max_issue}
 \end{figure}

We note that a degree of conditioning was required to produce the values of $kH/2$ shown in Fig. \ref{fig:AK_local_global_steepness}(b). The values of $kH/2$ shown as solid grey dots are the maximum values observed within $\pm T_c/2$ of the time of maximum slope $|\mathrm{min}(\eta_x)|$. The open markers show the unconstrained maximum values of $kH/2$ measured at any time during each simulation, which show considerably more scatter. Estimating wavenumber using zero-crossings is unstable and can cause large spikes in corresponding values of $kH/2$. This is further illustrated in Fig. \ref{fig:HK_max_issue}, where in panel (a) up- and down-crossing steepness of the largest wave is plotted as a function of time alongside maximum surface slope $|\mathrm{min}(\eta_x)|$. The surface elevations that correspond to the maximum values of each parameter, indicated by the vertical dashed lines in Fig. \ref{fig:HK_max_issue}(a), are plotted in Fig. \ref{fig:HK_max_issue}(b) with zero crossings shown as $\times$ symbols. Spikes in $kH/2$ correspond to instances where the free surface forms a wave with two zero crossings within close proximity, leading to a very short wavelength and large $k$.

Locally measured maximum surface slope $|\mathrm{min}(\eta_x)|$, shown in Fig. \ref{fig:AK_local_global_steepness} (c), does not vary significantly as a function of bandwidth for the largest non-breaking waves (large grey dots). This may suggest that, independent of bandwidth, breaking is triggered by a maximum value of local slope, which is consistent with the rational presented in \cite{pizzo2019a}.

The values of $S$ for breaking waves are all slightly larger than those for the largest non-breaking waves (Fig. \ref{fig:AK_local_global_steepness} (a)); this expected as $S$ is the independent variable we varied in our search for the breaking threshold. Some values of local steepness $kH/2$ are smaller for breaking than they are for non-breaking waves (Fig. \ref{fig:AK_local_global_steepness} (b)), making it a less useful threshold parameter. Finally, for local slope $\eta_x$ it appears that there may be a threshold value of approximately $0.55-0.60$ after which breaking occurs, causing values of local slope to increase sharply as the surface overturns (Fig. \ref{fig:AK_local_global_steepness} (c)). Evidently, local slope $\eta_x\rightarrow-\infty$ as overturning occurs; the values of slope in Fig. \ref{fig:AK_local_global_steepness} (c) for breaking waves (small red dots) are taken one time step prior to overturning.
 
Observed maximum local slope $\eta_x$ may be affected by the resolution of our simulations. Moreover, the onset of breaking may also change subtly depending on particle spacing at the wave crest. To establish the sensitivity of our results in Fig. \ref{fig:AK_local_global_steepness}(c) to the resolution of our numerical model, we re-run simulations of the largest non-breaking waves using the conformal mapping procedure outlined in \S\ref{sec:Num_ICs}. The results of these simulations are shown as smaller dark grey dots in Fig. \ref{fig:AK_local_global_steepness}(c). For these simulations with significantly increased resolution near the wave crest, local slope still appears to approach a limit. The horizontal dashed line in Fig. \ref{fig:AK_local_global_steepness}(c) corresponds to a value of to $1/\tan(\pi/3)$ ($\approx0.5774$), which corresponds to a slope of $60^\circ$ from the vertical or an enclosed crest angle of $120^\circ$ (i.e, the limiting waveform of \cite{stokes1880}). 

Simulations of wave groups with larger bandwidths will involve more waves of high frequency and short wavelength, and the modelling of these short waves will be affected more by the resolution of our simulations ($N_p$). Thus, it follows that the differences observed between simulations performed using Configurations A and B are likely to be an increasing function of bandwidth. This effect may also be the cause of the slight downward trend of $|\mathrm{max}(\eta_x)|$ with $\nu$ observed in Fig. \ref{fig:AK_local_global_steepness}(c) for the simulations performed using Configuration A. 

\begin{figure}
 	\centering
 	\includegraphics[scale=0.6]{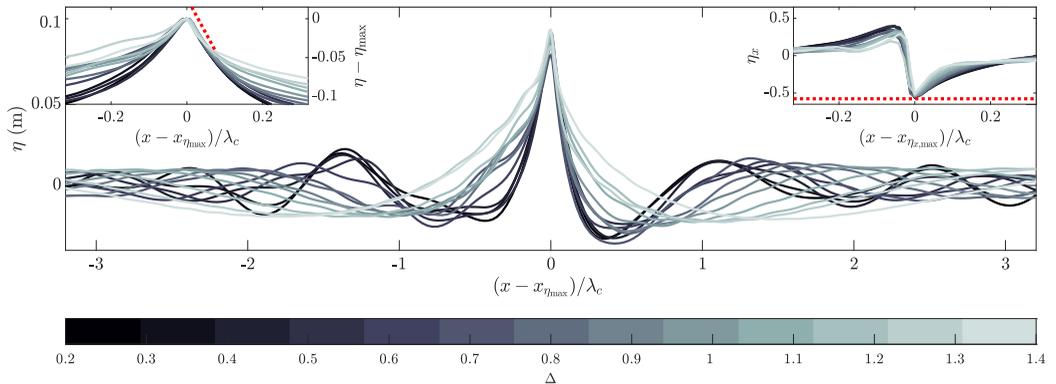}
 	\caption{Surface elevation of maximally steep non-breaking focused wave groups based on constant-steepness spectra of different bandwidths, plotted as a function of $x$, which has been shifted to align the position of maximum slope $\eta_x$ for different bandwidths and made non-dimensional using $\lambda_c$. Inset plots show vertically aligned surface elevation $\eta-\eta_\mathrm{max}$ (left) and local surface slope $\eta_x$ (right) at the wave crest. Line colours, dark to light, correspond to the bandwidth of the underlying spectrum, which ranges from $\Delta=0.2$ to $1.4$. The red dotted line has a slope of $1/\tan(\pi/3)$ ($\approx0.5774$), which corresponds to the limiting waveform of \cite{stokes1880}.}
 	\label{fig:eta_AK_thresh}
 \end{figure}
Surface elevation for the steepest simulated non-breaking waves based on constant-steepness spectra are plotted at the time of maximum local slope in Fig. \ref{fig:eta_AK_thresh}. The horizontal axis for each plot has been shifted so that the wave crests are aligned in space. The inset plots show the vertically aligned surface elevations and local slope. The lines in Fig. \ref{fig:eta_AK_thresh} correspond to the grey markers in Fig. \ref{fig:AK_local_global_steepness}, and are colored corresponding to their input bandwidth (increasing $\Delta$ dark to light).
This figure visualises the phenomena behind what we observe in Fig. \ref{fig:AK_local_global_steepness} (b) and (c). As bandwidth is increased, vertical asymmetry and zero-crossing wavelength of the waves both increase (i.e., $k\downarrow$ and $H\downarrow$), meaning that the overall waveforms appear quite different. However, the local (downward) slope of the wave crest front ($\eta_x<0$) across all bandwidths appear very similar; this is most evident in the inset plots of surface elevation and slope in Fig. \ref{fig:eta_AK_thresh}.
\subsection{Crest instability}
\begin{figure}
 	\centering
 	\includegraphics[scale=0.6]{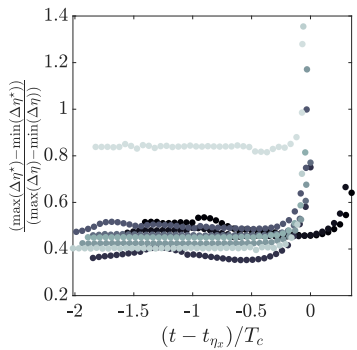}
 	\caption{Growth of a normalised surface perturbation as a function of time relative to the time $t_{\eta_x}$ at which surface slope reaches a value of $-0.5774$ for wave groups based on constant-steepness spectra. The different colored markers, dark to light, correspond to the bandwidth of the underlying spectrum, which ranges from $\Delta=0.2$ to $1.4$.}
 	\label{fig:instability_AK}
 \end{figure}

If breaking is indeed triggered by the local slope reaching a critical value, this may indicate that breaking is caused by a form of super-harmonic instability \citep{longuet1997}. Crest instabilities can occur when the surface slope reaches a critical value; this forms the basis of the breaking model in \cite{pizzo2019a}. To examine if such an instability is the cause of breaking in our simulations we compare the surface elevation of breaking and non-breaking waves for a given bandwidth \citep[see also][]{longuethiggins1978}. Specifically, we subtract the surface elevation of the largest non-breaking wave group $\eta^\star_{-\Delta S}$ from the surface elevation of the smallest breaking wave group $\eta^\star$ to calculate the free surface perturbation: $\Delta\eta^\star=\eta^\star-\eta^\star_{-\Delta S}$ immediately prior to breaking. We do so for the wave groups we simulate based on constant-steepness spectra. If the free surface perturbation grows rapidly (e.g., exponentially), this indicates the presence of a crest instability. We note that these two wave groups (i.e., $\eta^\star$ and $\eta^\star_{-\Delta S}$) have a small difference in input steepness $\Delta S=5\times10^{-5}$; this small difference in the initial conditions of the two wave groups $\Delta S$ may cause variation in $\Delta\eta^\star$ that is not a result of a crest instability. To quantify which part of  $\Delta\eta^\star$ is simply a result of different initial steepness, we also calculate the free surface perturbation between the largest non-breaking wave $\eta^\star_{-\Delta S}$ and a wave group $\eta^\star_{-2\Delta S}$ with initial steepness $S$ which is $10\times10^{-5}$ smaller again: $\Delta\eta=\eta^\star_{-\Delta S}-\eta^\star_{-2\Delta S}$. Following \cite{longuet1997}, we calculate the growth of potential instabilities from the maximum height difference at each instance in time, $\mathrm{max}(\Delta\eta^\star)-\mathrm{min}(\Delta\eta^\star)$, of the free surface perturbation. In Fig. \ref{fig:instability_AK} we plot $\mathrm{max}(\Delta\eta^\star)-\mathrm{min}(\Delta\eta^\star)$ normalised by $\mathrm{max}(\Delta\eta)-\mathrm{min}(\Delta\eta)$ as a function of time, where $t_{\eta_x}$ is the time at which surface slope reaches a value of $0.5774$. The normalised perturbation grows sharply around $t_{\eta_x}$ and is approximately constant prior to $t_{\eta_x}$. This sudden increase is indicative of unstable behavior, and suggest that surface slope may be subject to a type of (local) super-harmonic instability.   

\subsection{Spectral shape}\label{sec:spec_shape}
Linear calculations performed in \S \ref{sec:Lin_kin_Steep} demonstrated that, alongside bandwidth, spectral shape has a significant influence on surface kinematics and steepness of focused wave groups. In the following section we perform simulations of steep breaking and non-breaking focused wave groups to determine how spectral shape affects the relationship between breaking onset and bandwidth. As in \S \ref{sec:Lin_kin_Steep} we analyse simulations of focused wave groups based on constant-steepness, constant-amplitude, and JONSWAP spectra. We consider global steepness (\S \ref{sec:Global steepness}) and local slope (\S \ref{Local slope}) in turn. 
\subsubsection{Global steepness}
\label{sec:Global steepness}

\begin{figure}
 	\centering
 	\includegraphics[scale=0.6]{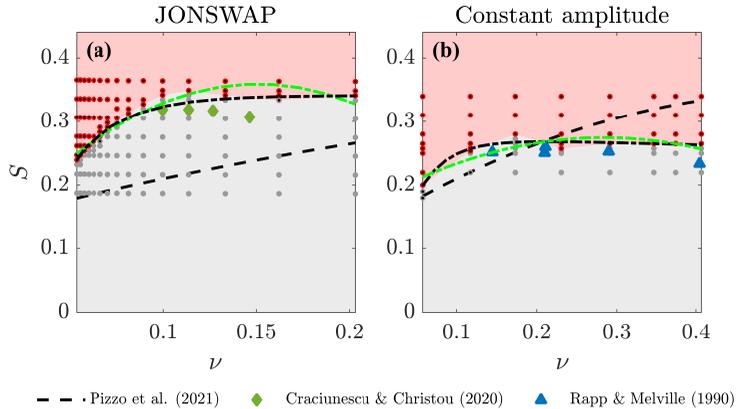}
 	\caption{Regime diagram of breaking and non-breaking behaviour as a function of global steepness $S$ and bandwidth $\nu$ for wave groups based on JONSWAP (a) and constant-amplitude (b) spectra. Grey markers indicate no breaking, red markers overturning breaking, and small black dots indicate $B>0.855$. Additional colored markers show comparable experimental (albeit finite-depth) results from \cite{rapp1990} and \cite{craciunescu2020}. Dot-dashed lines show our exponential (black) and quadratic (green) parametric curves fitted to the breaking onset steepness (see Tab. \ref{tab:fit_coeffs}). }
 	\label{fig:Phase_space_spec_shape}
\end{figure}
 \begin{table}
	\caption{Coefficients of parametric curves for breaking onset steepness as a function of bandwidth $\nu$ for focused wave groups based on JONSWAP and constant-amplitude spectra (see Fig. \ref{fig:Phase_space_spec_shape}).}\label{tab:fit_coeffs}
	\begin{center}
		\begin{tabular}{ccccc}
                \hline
			 &\multicolumn{4}{c}{$S=E_1\exp(E_2\nu)+E_3\exp(E_4\nu)$} \\
                \hline
			  & $E_1$ & $E_2$ & $E_3$ & $E_4$\\
                JONSWAP & $0.4026$&$-0.8732$&$-0.7611$&$-30.96$\\
                Constant amplitude & $0.2763$ &$-0.1196$&$-0.3502$&$-26.76$\\
			\hline  
                 & \multicolumn{3}{c}{$S=p_1 \nu^2+p_2 \nu+p_3$}&\\
                 \hline 
                    & $p_1$ & $p_2$ & $p_3$ & \\
                JONSWAP & $-11.5$&$3.4682$&$0.09799$&\\
                Constant amplitude & $-1.203$ &$0.6838$&$0.1775$&\\
                \hline
                \cite{pizzo2021} & \multicolumn{3}{c}{$S=-0.6948\nu^2+0.7541\nu+0.1417$}&\\ 
		\end{tabular}
	\end{center}
\end{table}

Figure \ref{fig:Phase_space_spec_shape} shows a regime diagram of breaking and non-breaking behaviour based on global steepness $S$ and bandwidth $\nu$ for focused wave groups based on the different spectra we simulate. Grey markers denote simulations where no breaking was observed, red markers denote simulations where breaking was detected, and black dots denote simulations where $B_x>0.855$ prior to overturning. The dashed black lines show \eqref{eq:Thresh_Pizz} obtained by \cite{pizzo2021} for a constant-steepness spectrum.

Before we discuss these simulations we emphasise that the range of bandwidths over which we simulate wave groups for each type of spectrum is different; thus the horizontal axes limits for each panel in Figs. \ref{fig:Phase_space_spec_shape} and \ref{fig:Phase_space_spec_shape_nx} are markedly different. For JONSWAP spectra varying $\gamma$ from 1 to 15 leads to variation in $\nu$ of approximately $0.25$-$0.375$. Further increases in $\gamma$ leads to minimal reduction in $\nu$. 
For constant-steepness and constant-amplitude spectra we are able to vary the bandwidth over a similar range. 

In Fig. \ref{fig:Phase_space_spec_shape}, the breaking threshold in terms of global steepness $S$ for focused wave groups based on JONSWAP spectra increases with bandwidth, but does not follow \eqref{eq:Thresh_Pizz}. For wave groups based on constant-amplitude spectra, initially there is a slight increase then a leveling-off of the breaking threshold $S$ with bandwidth. Both spectral shapes, JONSWAP and constant-amplitude, show similar behaviour, leveling off for large values of $\nu$, but at two different values, $\nu=0.34$ and $0.26$, respectively. Tab. \ref{tab:fit_coeffs} shows parametric curves we have obtained by fitting two different functional forms to the data for the different spectra in Fig. \ref{fig:Phase_space_spec_shape}.

\begin{figure}
 	\centering
 	\includegraphics[scale=0.6]{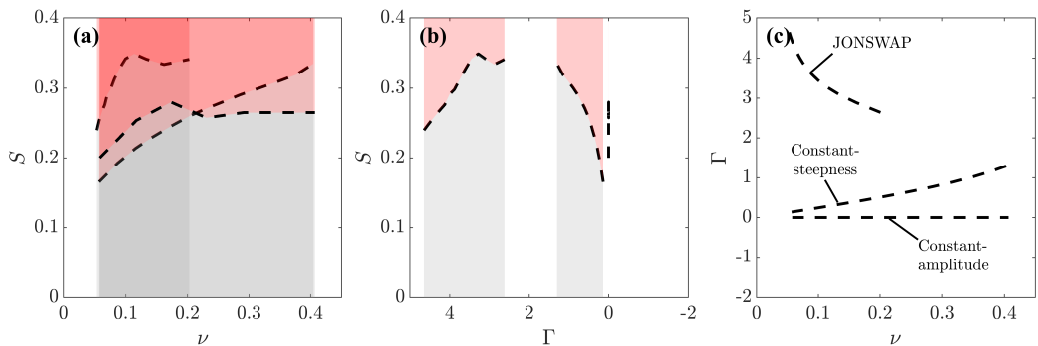}
 	\caption{Two-dimensional projections of the three-dimensional regime diagram of breaking and non-breaking behaviour in terms of global steepness $S$, skewness $\Gamma$, bandwidth $\nu$, showing breaking (red) and non-breaking (grey) behaviour for all the focused wave groups simulated. Vertical planes show the limits of the parameter range for each spectral shape (see labels in panel (c)). The breaking threshold is delineated using black dashed lines. All panels show different views of the the same plot.}
 	\label{fig:Phase_space_3D}
\end{figure}

As the next order spectral moment from bandwidth, spectral skewness may provide a way to characterise the influence spectral shape has on breaking onset. 
Spectra with a high-frequency tail (i.e., JONSWAP) will have positive skewness, as do constant-steepness spectra. For a constant-amplitude spectrum, skewness is zero, which may help explain the reduced variation of $S$ observed in Fig. \ref{fig:Phase_space_spec_shape}(b). In Fig. \ref{fig:Phase_space_3D} we plot two-dimensional projections of the regime diagram of breaking and non-breaking behaviour for all the simulations we perform in terms of bandwidth $\nu$, skewness $\Gamma$, calculated as the standardised third central moment of the energy spectrum, and global steepness $S$. The black dashed lines delineate breaking and non-breaking behaviour for each of the three types of spectra simulated. From the limited coverage of the parameter space that our simulations provide, it does not appear possible to fit a smooth surface to delineate breaking and non-breaking behaviour. 

\subsubsection{Local slope}
\label{Local slope}

\begin{figure}
 	\centering
 	\includegraphics[scale=0.6]{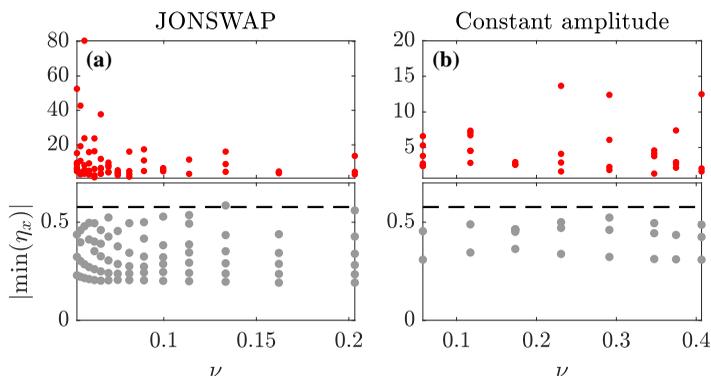}
 	\caption{Maximum local slope $\eta_x$ as a function of bandwidth for (a) JONSWAP and (b) constant-amplitude spectra. Red and grey markers correspond to simulations where breaking has and has not occurred, respectively. For breaking simulations $\eta_x$ was measured one time step prior to overturning. The horizontal dashed lines correspond to $1/\tan(\pi/3)$, i.e., a slope of $60^\circ$ or an enclosed crest angle of $120^\circ$. }
 	\label{fig:Phase_space_spec_shape_nx}
\end{figure}

In Fig. \ref{fig:Phase_space_spec_shape_nx} we plot maximum values of local slope $\eta_x$ measured during the numerical simulations of wave groups based on JONSWAP and constant-amplitude spectra. As observed for focused wave groups based on constant-steepness spectra, maximum local steepness for non-breaking waves appears to also approach the limit $1/\tan(\pi/3)$, i.e., a slope of $60^\circ$. This suggests that a breaking threshold based on local slope may be more universal than one based on global steepness, being valid across a range of bandwidths and spectral shapes. 

\subsection{Breaking onset detection}\label{sec:onset}
Above, we have found that the relationship between spectral shape and wave breaking onset is too complex to parameterise in a simple manner using a spectral parameters (i.e., $\nu$ and $\Gamma$) and global steepness $S$. Nevertheless, local parameters may still be used to indicate when breaking will occur on a wave-by-wave basis.  While less generally applicable and less predictive, readily observable local parameters that can detect breaking onset are still highly useful.

As previously mentioned, the parameter $B_x$ has been shown to be a promising means by which to detect the onset of breaking \citep{saket2017,barthelemy2018,derakhti2020}. In Figs. \ref{fig:AK_local_global_steepness} and \ref{fig:Phase_space_spec_shape_nx} we have shown that local maximum surface slope, defined as $|\mathrm{min}(\eta_x)|$, may approach a threshold prior to breaking that is independent of spectral shape and bandwidth. It appears that local surface slope at the wave crest $|\mathrm{min}(\eta_x)|\rightarrow 1/\tan(\pi/3)$ and may be self-similar for maximally steep waves (cf. Fig. \ref{fig:eta_AK_thresh}). This local maximum slope is the same maximum slope predicted by \cite{stokes1880} for the limiting waveform of progressive waves on deep water. Stokes's prediction of a $120^\circ$ corner flow 
at the crest of the limiting waveform, corresponds to a kinematic limit where $u/C=1$. Thus, there may be an inherent link between $|\mathrm{min}(\eta_x)|$ and $B_x$.

\begin{figure}
 	\centering
 	\includegraphics[scale=0.6]{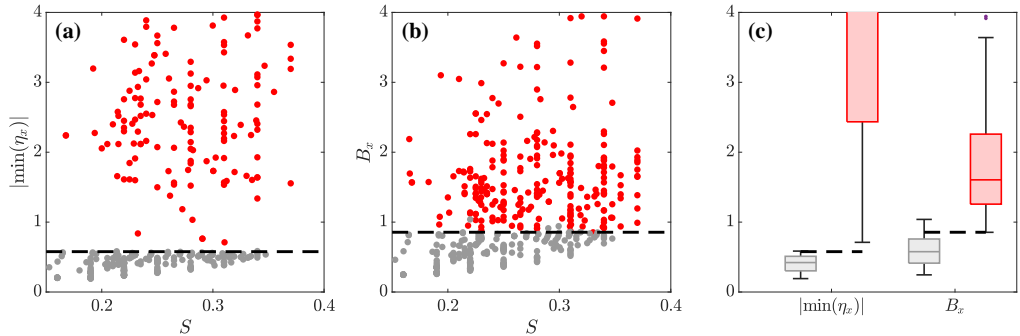}
 	\caption{Breaking onset threshold behaviour for all simulated wave groups, (a) maximum local slope $|\mathrm{min}(\eta_x)|$ and (b) parameter $B_x$ plotted as a function input global steepness; in panel (c) box and whiskers show the range of breaking breaking and non-breaking values for each parameter. Breaking and non-breaking waves are denoted by grey and red lines respectively. Black dashed lines show values of  $|\mathrm{min}(\eta_x)|=1/\tan(\pi/3)$ and $B_x=0.855$. }
 	\label{fig:etax_bx_comp}
\end{figure}

In many applications, particularly in the field or the laboratory, measuring local slope may be more straightforward than measuring surface kinematics. Therefore, $|\mathrm{min}(\eta_x)|$ may provide an alternative means to $B_x$, to detect the onset of breaking in such scenarios. In Fig. \ref{fig:etax_bx_comp} we plot breaking (red) and non-breaking (grey) values of $|\mathrm{min}(\eta_x)|$ (panel a) and $B_x$ (panel b) as a function of input global steepness $S$ for all the simulations we have performed. Threshold values of $|\mathrm{min}(\eta_x)|=1/\tan(\pi/3)$ and $B_x=0.855$ are shown as black dashed lines. In panel (c) box plots show the range of values both parameters take for breaking and non-breaking simulations. The top and bottom of each box correspond to the 25th and 75th percentiles, the central marks corresponds to the median, and the whiskers correspond to the range of values. Our simulations show that $|\mathrm{min}(\eta_x)|$ may function well as a threshold parameter as the separation between breaking and non-breaking is large and indeed larger than for $B_x$. The range of $|\mathrm{min}(\eta_x)|$ also does not extend above the limit $|\mathrm{min}(\eta_x)|=1/\tan(\pi/3)$. 
\section{Conclusions}\label{sec:conclusions}
Existing studies of wave breaking present a range of sometimes conflicting conclusions on how spectral bandwidth affects the onset of wave breaking; in many of these studies studying the effect of bandwidth on breaking onset may not have been the primary focus. \cite{pizzo2019a} and \cite{pizzo2021} directly addressed this question analytically and numerically and found that breaking onset occurs at increasing steepness as bandwidth is increased. \cite{wu2004} addressed the same problem experimentally and seemingly drew the opposite conclusion. Herein, we have performed linear and fully nonlinear simulations of two-dimensional focused wave groups to elucidate the potential causes of these differences, and gain a broader understanding of how bandwidth and spectral shape affect wave breaking onset. 
Linear calculations show that, on a kinematic basis alone, bandwidth will cause the onset of breaking to occur at lower global steepness $S$. For a constant global steepness $S$, bandwidth can cause apparent local steepness $kH/2$ to reduce by as much as $80\%$. Thus, even on a linear basis, the different measures of steepness used may serve to explain the differences in perceived breaking thresholds as a function of bandwidth in the existing literature. 

For fully nonlinear simulations of maximally steep focused wave groups based on constant-steepness spectra of varying bandwidth, we reproduce the breaking threshold in \cite{pizzo2021} and we obtain qualitatively similar results to \cite{wu2004}. Our numerical results demonstrate further that the definition of steepness is likely the main cause of the difference between these two studies. Additionally, we find that local steepness $kH/2$ is not a robust parameter because of rapid variation in zero-crossing wavelength and does not demarcate breaking and non-breaking wave groups effectively. 

For wave groups based on JONSWAP and constant-amplitude spectra, breaking onset steepness $S$ increases with bandwidth. Variation in breaking onset steepness is minimal for JONSWAP spectra of $\gamma=1$ to $5$ ($\nu=0.2$ to $0.1$), and all but the most narrow-banded constant-amplitude spectra wave groups. Global steepness $S$ demarcates breaking and non-breaking behavior as a function of bandwidth $\nu$ for all the spectral shapes we examine. Alongside bandwidth, we demonstrate that spectral shape affects the onset of wave breaking. However, we find that skewness and bandwidth alone do not provide enough information to parameterise the effects spectral shape has on breaking onset when considering global steepness $S$. 

It may be possible use additional spectral moments to try and parameterise breaking onset; though this may not be necessary in practise as for JONSWAP spectra, which best reflect ocean wave conditions, there is very little variation in breaking onset steepness $S$ at values of $\gamma=1-5$. Therefore, for the majority of sea states there may not be a great deal of variation in breaking onset over the range of realistic bandwidths. 

In contrast to local steepness $kH/2$ and global steepness $S$, the local slope $\eta_x$ of maximally steep non-breaking waves varies little as a function of bandwidth and approaches a threshold of $1/\tan(\pi/3)\approx 0.58$. This wave breaking threshold based on local slope is observed for all of the wave groups we examine regardless of spectral shape. 

Concluding, we have found that breaking onset occurs at the same threshold value of the local slope ($\partial \eta/\partial x$) regardless of the underlying spectrum of the waves. This is a remarkable result, which we do not believe has been reported before. Our simulations have demonstrated that $|\mathrm{min}(\eta_x)|$ may provide an alternative breaking onset threshold to the ratio of fluid and crest speed, $B_x=u/C$; the threshold $|\mathrm{min}(\eta_x)|$ may be easier to apply in certain scenarios, for example when fluid velocities are unknown. Similarity in the crest shape of breaking waves may be linked to a kinematic description of breaking (and thus to a given value of $B_x$) and should be examined further.
 \section*{Declaration of Interests}
 The authors report no conflict of interest.

\bibliographystyle{abbrvnat}
\bibliography{Breaking_2D}
\end{document}